\documentclass[a4paper,12pt]{article}
\usepackage{amsmath,amsfonts,amssymb,amscd,enumerate}

\def\a{\alpha}
\def\p{\partial}
\def\A{\mathcal{A}}
\def\R{\mathbb{R}}
\def\C{\mathbb{C}}
\def\N{\mathbb{N}}
\def\g{\mathfrak{g}}

\begin{document}

\title{Noncommutative Differential Forms on the kappa-deformed Space}

\author{Stjepan Meljanac$^1$ and Sa\v{s}a Kre\v{s}i\'{c}-Juri\'{c}$^2$}
\date{}

\maketitle

\vskip 0.5cm
\noindent $^1$ Rudjer Bo\v{s}kovi\'{c} Institute, Bijeni\v{c}ka cesta b.b., 10000 Zagreb, Croatia\\
\noindent $^2$ Faculty of Natural and Mathematical Sciences, University of Split, Teslina 12,\\ 21000 Split, Croatia

\begin{abstract}
We construct a differential algebra of forms on the kappa-deformed space. For a
given realization of noncommutative coordinates as formal power series in the Weyl algebra
we find an infinite family of one-forms and nilpotent exterior derivatives. We derive
explicit expressions for the exterior derivative and one-forms in covariant and noncovariant
realizations. We also introduce higher-order forms and show that the
exterior derivative satisfies the graded Leibniz rule. The differential forms are generally not
graded-commutative, but they satisfy the graded Jacobi identity. We also consider the star-product
of classical differential forms. The star-product is well-defined if the commutator between the
noncommutative coordinates and one-forms is closed in the space of one-forms alone.
In addition, we show that in certain realizations the exterior derivative acting
on the star-product satisfies the undeformed Leibniz rule.
\end{abstract}

\vskip 0.5cm
\noindent{PACS numbers: 02.20.Sv, 02.20.Uw, 02.40.Gh}

\section{Introduction}

Recent years have witnessed a growing interest in the formulation of physical theories on noncommutative (NC) spaces. The structure of NC
spaces and their physical implications were studied in \cite{Doplicher}-\cite{Chaichian}. Such spaces have roots in quantum
mechanics where the canonical phase space becomes noncommutative (see \cite{Phys-NC-World} for a historical treatment and the references
therein). Classification of the NC spaces and investigation of their properties, in particular the development of a general theory suitable for
physical applications, is an important problem. In this note we investigate differential calculus in the Euclidean kappa-deformed
space. The kappa-space is a mild deformation of the Euclidean space whose coordinates $\hat x_\mu$, $\mu=1,2,\ldots ,n$,
satisfy a Lie algebra type commutation relations. The commutation relations for $\hat x_\mu$ depend on a deformation vector
$a\in \R^n$ which is on a very small length scale and yields the undeformed space when $a\to 0$. The kappa-space was studied by different groups,
from both the mathematical and physical points of view \cite{Lukierski-1}-\cite{Freidel}. It provides a framework for doubly
special relativity \cite{Amelino-Camelia-1}, \cite{Kowalski-Glikman}, and it has applications in quantum gravity \cite{Amelino-Camelia-4}
and quantum field theory \cite{Daskiewicz-1},  \cite{Daskiewicz-2}

A crucial tool in the development of a physical theory is differential calculus. There have been several attempts to develop
differential calculus in the kappa-deformed space \cite{Sitarz}, \cite{Dimitrijevic-3}. For a general associative algebra
Landi gave a construction of a differential algebra of forms in \cite{Landi}. In this work we present a construction of
differential forms and exterior derivative in the kappa-deformed space using realizations of the NC coordinates $\hat x_\mu$
as formal power series in the Weyl algebra. Our approach is based on the methods developed for algebras of
deformed oscillators and the corresponding creation and annihilation operators \cite{Wess}-\cite{Kempf}.
The realizations of the NC coordinates $\hat x_\mu$ in various orderings have been
found in \cite{Meljanac-1} and \cite{Meljanac-2}. The realization of a general Lie algebra type NC space in the
symmetric Weyl ordering has been given in \cite{Durov}.

The outline of the paper is as follows. In section 2 we present a novel construction of a differential algebra of forms
on the kappa-deformed space. The exterior derivative $\hat d$ and one-forms $\xi_\mu$ are defined as formal power series in the
Lie superalgebra generated by commutative coordinates $x_\mu$, derivatives $\p_\mu$ and ordinary one-forms $dx_\mu$.
The number of one-forms $\xi_\mu$ is the same as the number of NC coordinates $\hat x_\mu$, and the results are valid for a
general deformation vector $a\in \R^n$. In the present work we do not require compatibility of the differential structure
with a kappa-deformed symmetry. This distinguishes our approach from \cite{Sitarz} where compatibility of the differential
calculus with the kappa-deformed symmetry group was considered. This compatibility requires that in addition to $\xi_\mu$
there is an extra one-form $\phi$. The realizations of $\hat d$ and $\xi_\mu$ are related to realizations of
$\hat x_\mu$ through a system of partial differential equations. We also define higher-order forms and show that
$\hat d$ is a nilpotent operator which satisfies the graded Leibniz rule. However, the differential forms are generally not graded
commutative. In the smooth limit when $a\to 0$ our theory reduces to classical results. In section 3 we analyze the exterior derivative
and one-forms in covariant realizations of the kappa-deformed space. We show that the algebra generated by $\hat x_\mu$
and $\xi_\mu$ generally does not close under the commutator bracket since $[\xi_\mu,\hat x_\nu]$ may involve an infinite
series in derivatives $\p_\mu$. We have derived a condition for the commutator $[\xi_\mu,\hat x_\nu]$ to be closed
and found realizations in which the condition holds.
A similar analysis was carried out by Dimitrijevi\'{c} et. al. in \cite{Dimitrijevic-3},
but our results are more general and in certain aspects different.
Section 4 deals with the differential algebra of forms in noncovariant realizations. We introduce a general
Ansatz for the exterior derivative and find the corresponding one-forms in the left, right and symmetric left-right realization.
In these realizations the commutator $[\xi_\mu,\hat x_\nu]$ is always closed in the space
of one-forms $\xi_\mu$ alone. In section 5 we present a novel construction of the star-product of (classical) differential forms.
The star-product depends on realizations of $\hat x_\mu$ and is well-defined if the commutator $[\xi_\mu,\hat x_\nu]$ is closed
in the space of one-forms $\xi_\mu$ alone. We show that for differential forms with constant coefficients the star-product
is undeformed and graded-commutative. However, this property does not hold for arbitrary forms. Also, we consider the
induced exterior derivative acting on the star-product of differential forms. A short conclusion is given in section 6.

\section{Differential Forms}

In this section we present a general construction of a differential algebra of forms in the Euclidean kappa-deformed
space. This construction is based on realizations of the NC coordinates $\hat x_\mu$ as formal power series in
the Weyl algebra introduced in \cite{Meljanac-1} and \cite{Meljanac-2}. We find that for a given realization of $\hat x_\mu$
there is an infinite family of exterior derivatives $\hat d$ and one-forms $\xi_\mu$ where $\xi_\mu$ are obtained
by the action of $\hat d$ on $\hat x_\mu$. This infinite family includes two
canonical types of $\hat d$ and $\xi_\mu$ whose realizations are studied in detail in the following sections.

The $n$-dimensional kappa-deformed space is a noncommutative space of Lie algebra type with generators
$\hat x_1, \hat x_2, \ldots ,\hat x_n$ satisfying the commutation relations
\begin{equation}\label{01}
[\hat x_\mu, \hat x_\nu] = i(a_\mu \hat x_\nu - a_\nu \hat x_\mu), \quad a_\mu \in \R.
\end{equation}
The vector $a\in \R^n$ describes the deformation of the $n$-dimensional Euclidean space.
The Lie algebra satisfying \eqref{01} will be denoted by $\mathfrak{g}$. The structure constants of $\mathfrak{g}$
are given by
\begin{equation}\label{02}
C_{\mu\nu\lambda} = a_\mu\, \delta_{\nu\lambda}-a_\nu\, \delta_{\mu\lambda}.
\end{equation}
Our construction of the differential calculus uses realizations of $\hat x_\mu$
as formal power series in the deformation parameter $a$ with coefficients in the Weyl algebra. The Weyl algebra is
generated by the operators $x_\mu$ and $\p_\mu$, $\mu=1,2,\ldots ,n$, satisfying $[x_\mu,x_\nu]=[\p_\mu,\p_\nu]=0$
and $[\p_\mu,x_\nu]=\delta_{\mu\nu}$. It has been shown in \cite{Meljanac-1} and \cite{Meljanac-2}
that there exist infinitely many realizations of $\hat x_\mu$ of the form
\begin{equation}\label{03}
\hat x_\mu = \sum_\a x_\a\, \phi_{\a\mu}(\p),
\end{equation}
where $\phi_{\a\mu}$ is a formal power series
\begin{equation}\label{55}
\phi_{\a\mu}(\p) = \delta_{\a\mu}+\sum_{|k|\geq 1} c_k\,a^{|k|}\, \p^k.
\end{equation}
We denote $\p^k = \p_1^{k_1}\p_2^{k_2}\ldots \p_n^{k_n}$ where $k$ is a multi-index of length $|k|=\sum_\mu k_\mu$.
In the limit as $a\to 0$ we have $\phi_{\a\mu}\to \delta_{\a\mu}$, whence $\hat x_\mu$ become the
commutative coordinates $x_\mu$. A representation \eqref{03} of the NC coordinates $\hat
x_\mu$ will be called a $\phi$-realization. The NC coordinates $\hat x_\mu$ and derivatives $\p_\mu$ generate
a deformed Heisenberg algebra satisfying
\begin{equation}\label{69}
[\p_\mu, \hat x_\nu] = \phi_{\mu\nu}(\p).
\end{equation}
We will assume that the matrix $[\phi_{\mu\nu}]$ is invertible, allowing us to express $x_\mu$ as
\begin{equation}\label{59}
x_\mu = \sum_\a \hat x_\a\, \phi^{-1}_{\a\mu}(\p),
\end{equation}
where $\phi^{-1}_{\a\mu}(\p)$ is also a formal power series of the type \eqref{55}. The existence of $\phi^{-1}_{\mu\nu}$ implies
that threre is a vector space isomorphism between the symmetric algebra generated by $x_\mu$, $\mu=1,2,\ldots ,n$, and
the enveloping algebra of $\g$. This isomorphism will be important in defining
the star-product discussed in section 5. With regard to the action of the rotation algebra $so(n)$ the
realizations of the kappa-space can be divided into covariant \cite{Meljanac-2} and noncovariant \cite{Meljanac-1}.
Both types of realizations will be used in the construction of differential forms in sections 3 and 4.

It is useful to introduce a unital associative algebra $\A$ over $\C$ generated by $x_\mu$, $\p_\mu$ and ordinary one-forms
$dx_\mu$, $1\leq \mu \leq n$, satisfying the additional relations $[dx_\mu,x_\nu]=[dx_\mu,\p_\nu]=0$ and $\{dx_\mu,dx_\nu\}=0$
where $\{\, ,\, \}$ denotes the anticommutator. A basis for $\A$ consists of the monomials
\begin{equation}\label{07}
x_1^{\a_1}\ldots x_n^{\a_n}\p_1^{\beta_1}\ldots \p_n^{\beta_n} dx_{\sigma_1}\ldots dx_{\sigma_p}
\end{equation}
where $\a_i,\beta_i\in \N_0$ and $1\leq \sigma_1<\sigma_2\ldots <\sigma_p\leq n$ for $p=1,2,\ldots ,n$. We define a
$\mathbb{Z}_2$-gradation of $\A$ by $\A=\A_0 \oplus \A_1$ where $\A_0$ and $\A_1$ are spanned by the monomials \eqref{07}
with $p$ even and odd, respectively. The algebra $\A$ is equipped with the graded commutator defined on homogeneous
elements by
\begin{equation}\label{63}
[[u,v]] = uv-(-1)^{|u|\, |v|} vu,
\end{equation}
where $|u|$ denotes the degree of $u$, ($|u|=0$ or $|u|=1$). The commutator \eqref{63} makes $\A$ into a Lie superalgebra,
and it satisfies the graded Jacobi identity
\begin{equation}\label{80}
(-1)^{|u|\, |w|}[[u,[[v,w]]\,]]+(-1)^{|v|\, |u|}[[v,[[w,u]]\,]]+(-1)^{|w|\,|v|}[[w,[[u,v]]\,]]=0.
\end{equation}

Recall that in the ordinary Euclidean space the exterior derivative is given by $d=\sum_\a dx_\a\, \p_\a$. It is a nilpotent operator,
$d^2=0$, satisfying the commutation relation $[d,x_\mu]=dx_\mu$. Our goal is to construct smooth deformations of $d$ and $dx_\mu$,
denoted $\hat d$ and $\xi_\mu$, $\mu=1,2,\ldots ,n$, which preserve the basic relation
\begin{equation}\label{09}
[\hat d,\hat x_\mu] =\xi_\mu.
\end{equation}
Let us assume that $\hat d$ and $\xi_\mu$ are represented by
\begin{equation}\label{08}
\xi_\mu = \sum_\a dx_\a h_{\a\mu}(\p) \quad \text{and}\quad \hat d=\sum_{\a,\, \beta} dx_\a \p_\beta k_{\a\beta}(\p),
\end{equation}
where $h_{\mu\nu}$ and $k_{\mu\nu}$ are formal power series of the type \eqref{55}. The boundary conditions
$\lim_{a\to 0}h_{\mu\nu} = \delta_{\mu\nu}$ and $\lim_{a\to 0}k_{\mu\nu} = \delta_{\mu\nu}$ ensure that in the smooth limit
$\xi_\mu \to dx_\mu$ and $\hat d\to d$ as $a\to 0$. As in the classical case, the deformed one-forms
anticommute and the exterior derivative is nilpotent. Indeed,
\begin{align}\label{62}
\{\xi_\mu,\xi_\nu\} &= \sum_{\alpha <\beta} \{dx_{\a},dx_{\beta}\} \left(h_{\a\mu}h_{\beta\nu}+h_{\a\nu}h_{\beta\mu}\right)=0, \\
\hat d^2 &= \sum_{\a<\beta} \{dx_\a, dx_{\beta}\} \sum_{\mu, \nu} \p_\mu\p_\nu k_{\a\mu} k_{\beta\nu} = 0,
\end{align}
since $\{dx_\a,dx_\beta\}=0$. We assume that the matrix $[h_{\mu\nu}]$ is invertible so that we may express $dx_\mu$
in terms of $\xi_\mu$. Using representation \eqref{08} one finds that the commutation relation \eqref{09}
is equivalent to a system of partial differential equations for the unknown functions $h_{\mu\nu}$ and $k_{\mu\nu}$:
\begin{equation}\label{10}
\sum_\rho \left(k_{\a\rho}+\sum_\beta \frac{\p k_{\a\beta}}{\p \p_\rho}\, \p_\beta \right)\phi_{\rho\mu} = h_{\a\mu}.
\end{equation}
This is an underdetermined system of $n^2$ equations for $2n^2$ unknown functions.
Taking the commutator of $\hat d$ with both sides of the commutation relations \eqref{01}, and applying the Jacobi
identity to the commutator $[\hat d,[\hat x_\mu,\hat x_\nu]]$, we find that $\hat x_\mu$ and $\xi_\nu$ satisfy the
compatibility condition
\begin{equation}\label{11}
[\hat x_\mu,\xi_\nu]-[\hat x_\nu,\xi_\mu]=i(a_\mu \xi_\nu-a_\nu \xi_\mu).
\end{equation}
Hence, every solution of Eq. \eqref{10} must be compatible with the differential equation implicit in \eqref{11}.
We note that Eq. \eqref{11} implies that, since $a\neq 0$, not all commutators $[\hat x_\mu,\xi_\nu]$ can be
simultaneously zero.

The condition \eqref{11} places constraints on the choice of $k_{\mu\nu}$ and $h_{\mu\nu}$. For a given function $k_{\mu\nu}$
satisfying $\lim_{a\to 0}k_{\mu\nu}=\delta_{\mu\nu}$, Eq. \eqref{10} uniquely determines $h_{\mu\nu}$. The boundary
conditions imposed on $\phi_{\mu\nu}$ and $k_{\mu\nu}$ imply that $\lim_{a\to 0}h_{\a\mu}=\delta_{\a\mu}$ automatically holds.
Therefore, starting with the exterior derivative $\hat d$ one readily finds the one-forms $\xi_\mu$ satisfying Eq. \eqref{09}.
However, the converse is not true since one cannot always find $k_{\mu\nu}$ for an arbitrary choice of $h_{\mu\nu}$. For example,
if $h_{\mu\nu}=\delta_{\mu\nu}$ then Eq. \eqref{08} implies that $\xi_\mu$ is the ordinary one-form, $\xi_\mu =dx_\mu$. In this
case $[\hat x_\mu,\xi_\nu]=0$ for all $\mu,\nu=1,2,\ldots n$, which contradicts the compatibility condition \eqref{11}.

Let $\bar \A$ denote the formal completion of $\A$.
We associate to the exterior derivative $\hat d$ a linear map or action $\hat d\colon \bar \A\to \bar \A$ defined by
\begin{equation}
\hat d\cdot u = [[\hat d,u]].
\end{equation}
It follows from Eq. \eqref{09} that $\hat d\cdot \hat x_\mu = \xi_\mu$, hence the action of $\hat d$ on the
coordinate $\hat x_\mu$ yields the one-form $\xi_\mu$. The action of $\hat d$ on the product of homogeneous elements
$u,v,\in\bar \A$ satisfies the graded Leibniz rule
\begin{equation}\label{64}
\hat d \cdot (uv) = (\hat d\cdot u)v +(-1)^{|u|}\, u (\hat d\cdot v).
\end{equation}
For zero-forms $\hat f = \hat f(\hat x)$ and $\hat g = \hat g(\hat x)$ this reduces to the undeformed Leibniz rule
\begin{equation}
\hat d \cdot (\hat f \hat g) = (\hat d \cdot \hat f)\hat g + \hat f (\hat d \cdot \hat g).
\end{equation}

It turns out that it is quite natural to consider the following canonical representation of $\hat d$ and $\xi_\mu$:\\

\noindent\textit{Type I}
\begin{equation}\label{42}
\hat d =\sum_\a dx_\a\, \p_\a, \quad \xi_\mu = \sum_\a dx_\a\, \phi_{\a\mu}(\p),
\end{equation}
\noindent\textit{Type II}
\begin{equation}\label{43}
\hat d = \sum_\a \xi_\a\, \p_\a, \quad \xi_\mu = \sum_\a dx_\a\, h_{\a\mu}(\p).
\end{equation}
The first type is obtained by choosing $k_{\mu\nu}=\delta_{\mu\nu}$, in which case Eq. \eqref{10} yields $h_{\mu\nu}=\phi_{\mu\nu}$.
This provides the simplest possible realization of the one-form $\xi_\mu$.
The second type is obtained by demanding that $k_{\mu\nu}=h_{\mu\nu}$. Then the functions
$h_{\mu\nu}$ satisfy the system of partial differential equations
\begin{equation}\label{15}
\sum_\rho \left(h_{\a\rho}+\sum_\beta \frac{\p h_{\a\beta}}{\p \p_\rho}\, \p_\beta\right) \phi_{\rho\mu} = h_{\a\mu}
\end{equation}
subject to the boundary conditions $\lim_{a\to 0}h_{\mu\nu}=\delta_{\mu\nu}$. In this case both the exterior derivative $\hat d$
and one-forms $\xi_\mu$ depend in a very nontrivial manner on the given $\phi$-realization. In the following sections sections
we shall analyze $\hat d$ and $\xi_\mu$ in covariant and noncovariant realizations found in \cite{Meljanac-1} and \cite{Meljanac-2}.
Note that the generators $\hat x_\mu, \p_\mu, \xi_\mu$, $\;1\leq \mu \leq n$, form an associative superalgebra which inherits
the grading from the superalgebra $\mathcal{A}$. The subalgebra generated by $\hat x_\mu, \p_\mu$, $1\leq \mu \leq n$ is
the deformed Heisenberg algebra \eqref{69}.

So far we have defined the exterior derivative $\hat d$ and one-forms $\xi_\mu$ such that $\hat d\cdot \hat x_\mu = \xi_\mu$.
We would like to extend the above construction to higher-order forms so that the action of $\hat d$ on $k$-forms yields
$(k+1)$-forms. First, we need to define what is meant by a $k$-form for $k\geq 1$. A $k$-form is a finite linear combination of monomials
in $\hat x_1, \hat x_2, \ldots ,\hat x_n$ and $\xi_1,\xi_2, \ldots ,\xi_n$ such that there are precisely $k$ one-forms $\xi_\mu$
in each monomial. The one-forms $\xi_\mu$ may be placed in any order in a given monomial. For example, both $\hat \omega^1=
\hat x_\mu \hat x_\nu \xi_\rho$ and $\hat \eta^1 = \hat x_\mu \xi_\rho \hat x_\nu$ are one-forms, albeit different. Let $\hat \Omega^k$
denote the space of $k$-forms and let $\hat \Omega = \bigoplus_{k\geq 0}\hat \Omega^k$.
The multiplication in $\hat \Omega$ is simply given by juxtaposition of the elements. This defines a grading on $\hat \Omega$ since
$\hat \Omega^k\, \hat \Omega^l \subseteq \hat \Omega^{k+l}$. We note that the product of differential forms is not graded-commutative
in general,
\begin{equation}
\hat \omega^k\, \hat \eta^l \neq (-1)^{kl}\, \hat \eta^l\, \hat \omega^k.
\end{equation}
The product is graded-commutative only for constant forms $\hat \omega^k = \xi_{\mu_1}\xi_{\mu_2}\ldots \xi_{\mu_k}$
since $\xi_{\mu_i}$ and $\xi_{\mu_j}$ anticommute.

Next we show that the exterior derivative $\hat d$ maps $\hat \Omega^k$ into $\hat \Omega^{k+1}$ for $k\geq 0$.
First, using the Leibniz rule \eqref{64} it is easily seen that
\begin{equation}\label{65}
\hat d\cdot \hat f(\hat x)\in \hat \Omega^1 \quad \text{for all}\quad \hat f(\hat x)\in \hat \Omega^0.
\end{equation}
Furthermore, using Eq. \eqref{08} we find
\begin{equation}
\hat d\cdot \xi_\mu = [[\hat d,\xi_\mu]] = \hat d\xi_\mu +\xi_\mu \hat d = 0
\end{equation}
since $\{dx_\mu,dx_\nu\}=0$. By induction on $k$ one can show that
\begin{equation}\label{66}
\hat d\cdot (\xi_{\mu_1}\xi_{\mu_2}\ldots \xi_{\mu_k})=0 \quad \text{for all}\quad k\geq 1.
\end{equation}
The relations \eqref{65} and \eqref{66} together with the Leibniz rule \eqref{64} imply that $\hat d$ maps
$k$-forms to $(k+1)$-forms. For example,
\begin{equation}
\hat d\cdot (\hat x_\mu \hat x_\nu \xi_\lambda) = \hat d\cdot (\hat x_\mu \hat x_\nu) \xi_\lambda
= \xi_\mu \hat x_\nu \xi_\lambda + \hat x_\mu \xi_\nu \xi_\lambda.
\end{equation}
The exterior derivative satisfies the graded Leibniz rule
\begin{equation}
\hat d\cdot \big(\hat \omega^k\, \hat \eta^l\big) = \big(\hat d\cdot \hat \omega^k\big) \hat \eta^l
+(-1)^k\, \hat \omega^k\, \big(\hat d\cdot \hat \eta^l\big).
\end{equation}
Hence, the algebra $\hat \Omega$ together with the linear map $\hat d\colon \hat \Omega^k\to
\hat \Omega^{k+1}$ is a differential algebra. Our approach is essentially the same as the construction of the differential
algebra of forms discussed in \cite{Landi}. In our case the algebra of zero-forms has the additional structure of the
universal enveloping algebra satisfying relations \eqref{01}. We note that in general one cannot rewrite a given $k$-form
such that $\xi_{\mu_1},\xi_{\mu_2},\ldots ,\xi_{\mu_k}$ are placed to the far right. This is possible only in special
realizations in which the commutator $[\xi_\mu, \hat x_\nu]$ closes in the space of one-forms $\xi_\mu$ alone.

\section{Covariant realizations}

In this section we shall investigate the differential algebra of forms in covariant realizations of the kappa-deformed
space introduced in \cite{Meljanac-2}. These realizations are covariant under the action of the rotation aglebra $so(n)$.
Of particular interest is a class of simple realizations obtained for the following choice of
$\phi_{\mu\nu}$ in the representation \eqref{03}:

\noindent{\textit{Left realization:}}
\begin{equation}
\phi_{\mu\nu} = (1-A)\delta_{\mu\nu}
\end{equation}

\noindent{\textit{Right realization:}}
\begin{equation}
\phi_{\mu\nu} = \delta_{\mu\nu} +ia_\nu \p_\mu
\end{equation}

\noindent{\textit{Natural realization:}}
\begin{equation}\label{20}
\phi_{\mu\nu}(\p)=(-A+\sqrt{1-B})\delta_{\mu\nu}+ia_\mu\p_\nu,
\end{equation}

\noindent{\textit{Symmetric realization:}}
\begin{equation}
\phi_{\mu\nu} = \frac{A}{e^A-1} \delta_{\mu\nu} + ia_\nu \p_\mu \frac{e^A-A-1}{(e^A-1)A}
\end{equation}
Here $A$ and $B$ are commuting operators defined by $A=ia\p$ and $B=a^2\p^2$ where we use the convention
$a\p = \sum_\a a_\a \p_\a$, $\p^2 = \sum_\a \p_\a^2$, etc. The symmetric realization corresponds to the
Weyl symmetric ordering of the monomials in $\hat x_\mu$. We remark that for a general Lie algebra type
NC space there is a universal formula for $\phi_{\mu\nu}$ in Weyl symmetric ordering given in \cite{Durov} as follows.
Suppose $\hat x_1,\hat x_2,\ldots ,\hat x_n$ are generators of a Lie algebra with structure constants $\theta_{\mu\nu\a}$:
\begin{equation}\label{71}
[\hat x_\mu,\hat x_\nu] = i\sum_\a \theta_{\mu\nu\a} \hat x_\a.
\end{equation}
Let $M=[M_{\mu\nu}]$ denote the $n\times n$ matrix of differential operators with elements
\begin{equation}
M_{\mu\nu} = i\sum_\a \theta_{\a\nu\mu} \p_\a.
\end{equation}
Then the Weyl symmetric realization of the Lie algebra \eqref{71} is given by
\begin{equation}
\phi_{\mu\nu}(\p)=p(M)_{\mu\nu}\quad \text{where}\quad  p(M)=\frac{M}{e^M-1}
\end{equation}
is the generating function for the Bernoulli numbers (see also \cite{Skoda}). In principle the exterior
derivative and one-forms may be constructed using any of the above realizations. Here we shall consider
the left, right and natural realization.

\subsection{Covariant realizations of type I}

Let us consider realizations of type I where the exterior derivative is undeformed,
$\hat d = \sum_\a dx_\a \p_\a$, and one-forms are given by $\xi_\mu =\sum_\a dx_\a \phi_{\a\mu}(\p)$.
We investigate the conditions under which the commutator $[\xi_\mu,\hat x_\nu]$ is closed in the space
of one-forms $\xi_\mu$. The closedness of the commuatator is important when considering the extended
star-product of (classical) forms in section 5.

Using realization \eqref{03} we have
\begin{equation}
[\xi_\mu,\hat x_\nu] = \sum_\a \sum_\beta dx_\a\, \frac{\p \phi_{\a\mu}}{\p \p_\beta}\, \phi_{\beta\nu}.
\end{equation}
The matrix $[\phi_{\mu\nu}]$ is invertible, hence we may express $dx_\mu$ in terms of $\xi_\mu$ to obtain
\begin{equation}\label{72}
[\xi_\mu, \hat x_\nu] = \sum_\sigma C_{\mu\nu\sigma}(\p) \xi_\sigma
\end{equation}
where
\begin{equation}
C_{\mu\nu\sigma}(\p)=\sum_\a \sum_\beta \phi^{-1}_{\sigma\a}\, \frac{\p \phi_{\a\mu}}{\p \p_\beta}\, \phi_{\beta\nu}.
\end{equation}
Clearly, the commutator \eqref{72} is closed in the space of one-forms $\xi_\mu$ only if the coefficients $C_{\mu\nu\sigma}$
are constant. This condition is satisfied in the left and right realizations, as shown in the following.
In the left realization we have
\begin{equation}
\hat x_\mu = x_\mu (1-A), \quad \xi_\mu = dx_\mu (1-A),
\end{equation}
which yields
\begin{equation}\label{73}
[\xi_\mu, \hat x_\nu] = -ia_\nu \xi_\mu.
\end{equation}
Similarly, in the right realization we have
\begin{equation}
\hat x_\mu = x_\mu +ia_\mu (x\p), \quad \xi_\mu = dx_\mu +ia_\mu (dx\p),
\end{equation}
which leads to
\begin{equation}\label{74}
[\xi_\mu, \hat x_\nu] = ia_\mu \xi_\nu.
\end{equation}
On the other hand, in the natural and symmetric realizations the coefficients $C_{\mu\nu\sigma}$ involve partial
derivatives so the commutators between $\xi_\mu$ and $\hat x_\nu$ are not closed.

\subsection{Covariant realizations of type II}

Consider now realizations of type II where the exterior derivative and one-forms are
given by $\hat d = \sum_\a \xi_\a \p_\a$ and $\xi_\mu = \sum_\a dx_\a h_{\a\mu}(\p)$, and $h_{\a\mu}$
is a solution of Eq. \eqref{15}. In this section we shall construct $\hat d$ and $\xi_\mu$ using the
natural realization \eqref{20}. The construction of NC forms in type II realization was considered in \cite{Dimitrijevic-3},
but not in a proper and complete way. Our motivation for using the natural realization is to present a proper analysis of this
problem.

Let us write Eq. \eqref{15} in a more compact form
\begin{equation}\label{17}
\sum_\rho \frac{\p \Lambda_\a}{\p \p_\rho}\, \phi_{\rho\mu} = h_{\a\mu}
\end{equation}
where $\Lambda_\a (\p)=\sum_\beta h_{\a\beta}(\p) \p_\beta$. The idea is to
solve an auxiliary problem for $\Lambda_\a$ and then calculate $h_{\mu\nu}$ from Eq. \eqref{17}.
Multiplying Eq. \eqref{17} by $\p_\mu$ and summing we obtain the following boundary value problem for $\Lambda_\a$:
\begin{equation}\label{18}
\sum_\rho \frac{\p \Lambda_\a}{\p \p_\rho} \Psi_\rho = \Lambda_a, \quad \lim_{a\to 0}\Lambda_\a = \p_\a,
\end{equation}
where $\Psi_\rho (\p) = \sum_\mu \phi_{\rho\mu}(\p)\p_\mu$. In the natural realization \eqref{20} we find
\begin{equation}\label{19}
\Psi_\rho (\p) = \p_\rho (-A+\sqrt{1-B}) +ia_\rho \p^2.
\end{equation}
Let us denote $Z^{-1}=-A+\sqrt{1-B}$. This is the inverse shift operator introduced in \cite{Meljanac-2}.
The index structure of $\Psi_\rho$ and Eq. \eqref{18}  suggest that we should look for $\Lambda_\a$ in the form
\begin{equation}\label{22}
\Lambda_\a (\p) = \p_\a H_1(A,B)+ia_\a \p^2 H_2(A,B)
\end{equation}
for unknown functions $H_1$ and $H_2$. From Eqs. \eqref{19} and \eqref{22} we obtain
\begin{align}
&\sum_\rho \frac{\p \Lambda_\a}{\p \p_\rho} \Psi_\rho =
\p_\a\left[\left(H_1+A\frac{\p H_1}{\p A}+2B\frac{\p H_1}{\p B}\right) Z^{-1}-B\frac{\p H_1}{\p A}+2AB\frac{\p H_1}{\p B}\right] \label{23} \\
&+ ia_\a \p^2\left[\left(2H_2+A\frac{\p H_2}{\p A}+2B\frac{\p H_2}{\p B}\right)Z^{-1}+H_1+2AH_2-B\frac{\p H_2}{\p A}+2AB\frac{\p H_2}{\p B}\right].
\notag
\end{align}
Substituting the above result into Eq. \eqref{18} we find that $H_1$ and $H_2$ satisfy the following system of differential equations:
\begin{align}
\Big(H_1+A\frac{\p H_1}{\p A}+2B\frac{\p H_1}{\p B}\Big) Z^{-1}-B\frac{\p H_1}{\p A}+2AB\frac{\p H_1}{\p B} &= H_1, \label{24} \\
\Big(2H_2+A\frac{\p H_2}{\p A}+2B\frac{\p H_2}{\p B}\Big) Z^{-1}-B \frac{\p H_2}{\p A}+2AB\frac{\p H_2}{\p B}+2AH_2+H_1 &= H_2. \label{25}
\end{align}
Since $\Lambda_\a (\p)\to \p_a$ as $a\to 0$, $H_1$ and $H_2$ are subject to the boundary conditions
\begin{equation}\label{67}
\lim_{a\to 0} H_1(A,B)=1, \quad \lim_{a\to 0}H_2(A,B) \;\; \text{finite}.
\end{equation}
It is shown in Appendix A that the above system has a unique solution
\begin{align}
H_1(A,B) &= \frac{2(1-\sqrt{1-B})}{B(-A+\sqrt{1-B})}, \\
H_2(A,B) &= -2(1-A+\sqrt{1-B})\left(\frac{1-\sqrt{1-B}}{B}\right)^2.
\end{align}
Inserting the expressions for $H_1$ and $H_2$ into Eq. \eqref{22} we find
\begin{equation}\label{33}
\Lambda_\a(\p) = \p_\a\, \frac{2(1-\sqrt{1-B})}{B(-A+\sqrt{1-B})} -ia_\a \p^2\, 2(1-A+\sqrt{1-B})\left(\frac{1-\sqrt{1-B}}{B}\right)^2.
\end{equation}
Since the exterior derivative is given by $\hat d = \sum_\a \xi_a\, \p_\a$ where $\xi_\mu =\sum_\a
dx_\a\, h_{\a\mu}(\p)$, $\hat d$ can be expressed in terms of $\Lambda_\a$ as
\begin{equation}
\hat d = \sum_\a dx_\a\, \Lambda_\a(\p).
\end{equation}
Thus, we find from Eq. \eqref{33} that
\begin{equation}
\hat d = \frac{2(1-\sqrt{1-B})}{B(-A+\sqrt{1-B})}\, (\p dx)-2(1-A+\sqrt{1-B})\left(\frac{1-\sqrt{1-B}}{B}\right)^2 i(adx)\p^2.
\end{equation}
Keeping only the first-order terms in $a\in \R^n$ we obtain the approximation
\begin{equation}
\hat d = \p dx +i(a\p) (\p dx)-i\p^2 (adx),
\end{equation}
where $d=\p\, dx$ is the undeformed exterior derivative.

Next we consider the one-form $\xi_\mu$. Substituting Eqs. \eqref{20} and \eqref{33} into Eq. \eqref{17} we find after some manipulation that
\begin{equation}
h_{\a\mu}(\p) = L_1 \delta_{\a\mu}+iL_2a_\a \p_\mu +iL_3a_\mu\p_a +a^2 L_4 \p_\a \p_\mu -\p^2 L_5 a_\a a_\mu, \label{41}
\end{equation}
where
\begin{align}
L_1 &= \frac{2(1-\sqrt{1-B})}{B},  \\
L_2 &= -\frac{2(-1+\sqrt{1-B})\left[2(A^2+A-B)\sqrt{1-B}+B-2(A^2-2AB+A)\right]}{B^2 (-A+\sqrt{1-B})}, \\
L_3 &= \frac{2(1-\sqrt{1-B})}{B(-A+\sqrt{1-B})}, \\
L_4 &= -\frac{2(B+2\sqrt{1-B}-2)}{B^2(-A+\sqrt{1-B})}, \\
L_5 &= \frac{2(-A+\sqrt{1-B})(1-\sqrt{1-B})^2}{B^2}.
\end{align}
Therefore, in the natural realization of type II the one-form $\xi_\mu$ is given by
\begin{align}
\xi_\mu &= \sum_\a h_{\a\mu}(\p)\, dx_\a \notag \\
&= L_1 dx_\mu +\left(iL_2\p_\mu -\p^2 L_5 a_\mu\right)(adx)+\left(iL_3a_\mu +a^2 L_4 \p_\mu\right)(\p dx). \label{37}
\end{align}
Although the above realization of $\xi_\mu$ is rather complicated, the first-order approximation has a particularly nice form
\begin{equation}\label{36}
\xi_\mu = dx_\mu +\sum_\a i(a_\mu \p_\a -a_\a \p_\mu) dx_\a.
\end{equation}

Let us now investigate the commutation relations for $\xi_\mu$ and $\hat x_\nu$. The NC coordinates in the natural realization
\eqref{20} are given by
\begin{equation}
\hat x_\mu = x_\mu (-A+\sqrt{1-B})+i(ax)\p_\mu.
\end{equation}
The explicit form of the commutator $[\xi_\mu,\hat x_\nu]$ is fairly complicated and a complete derivation is given in Appendix B.
Here we only state that it can be expressed as
\begin{equation}\label{40}
[\xi_\mu,\hat x_\nu] = \xi_\mu \frac{P_\nu^{(1)}}{L_1} + \xi_\nu\, \frac{P^{(2)}_\mu}{L_1} + (ia\xi)\, R^{(1)}_{\mu\nu} + (\p \xi)\, R^{(2)}_{\mu\nu}
\end{equation}
where $P^{(i)}_\mu$ and $R^{(i)}_{\mu\nu}$ are certain combinations of the functions $L_1,L_2,\ldots ,L_5$ and their partial derivatives.
We note that the commutator \eqref{40} is not closed since the right-hand side involves derivatives $\p_\mu$.
To gain an insight into the form of the commutator it is instructive to find a first-order approximation
in the parameter $a$. To first order in $a$ the natural realization of $\hat x_\mu$ is given by
\begin{equation}\label{38}
\hat x_\mu = x_\mu (1-ia\p)+i(ax)\p_\mu.
\end{equation}
Using the approximations \eqref{36} and \eqref{38} we obtain
\begin{equation}\label{39}
[\xi_\mu, \hat x_\nu] = i\sum_\a (a_\mu \delta_{\a\nu}-a_\a \delta_{\mu\nu})\xi_\a.
\end{equation}
As a special case suppose that the vector $a\in \R^n$ has only one non-zero component,
$a_\mu = a\delta_{\mu n}$ for $\mu=1,2,\ldots ,n$. Then
\begin{equation}
[\xi_\mu,\hat x_\nu] = ia(\delta_{\mu n}\, \xi_\nu - \delta_{\mu\nu}\, \xi_n).
\end{equation}
The above result agrees to first order in $a$ with the commutator $[\xi_\mu, \hat x_\nu]$ for vector-like transforming
one-forms considered in \cite{Dimitrijevic-3}. We emphasize, however, that the exact expression \eqref{40} does not agree with this
commutator for higher orders in $a$.

\section{Noncovariant realizations}

In this section we consider the exterior derivative and one-forms in noncovariant realizations of the
kappa-space introduced in \cite{Meljanac-1}. We assume that the components of the deformation vector $a\in \R^n$ are given by
$a_k=0$ for $k=1,2,\ldots, n-1$ and $a_n =a$. Then the commutation relations \eqref{01} yield
\begin{equation}\label{52}
[\hat x_k, \hat x_l]=0, \quad [\hat x_n, \hat x_k]=ia\hat x_k, \quad k,l=1,2,\ldots ,n-1.
\end{equation}
We use the Latin alphabet for the indices $1,2,\ldots,n-1$ and the Greek alphabet for the full set $1,2,\ldots, n$.
It was shown in \cite{Meljanac-1} that the NC coordinates $\hat x_\mu$ have infinitely many realizations of the form
\begin{align}
\hat x_k &= x_k\, \varphi(A), \quad k=1,2,\ldots, n-1,  \label{48} \\
\hat x_n &= x_n +ia \sum_{k=1}^{n-1} x_k \p_k\, \gamma(A),  \label{49}
\end{align}
where
\begin{equation}
\gamma(A)=\frac{\varphi^\prime (A)}{\varphi (A)}+1, \quad A=ia\p_n.
\end{equation}
The realizations are parametrized by the function $\varphi(A)$ satisfying the boundary conditions $\lim_{a\to 0}\varphi(A)=1$ and
$\lim_{a\to 0}\varphi^\prime (A)$ finite, so that $\hat x_\mu \to x_\mu$ as $a\to 0$.
The NC coordinates $\hat x_\mu$ are covariant under the rotation algebra $so(n-1)$, but not generally under the full
algebra $so(n)$.

The most general Ansatz for the exterior derivative $\hat d$ invariant under $so(n-1)$ is
\begin{equation}
\hat d = \sum_{k=1}^{n-1} dx_k\, \p_k\, N_1(A,\Delta)+dx_n\, \p_n\, N_2(A,\Delta)+ia\, dx_n \sum_{k=1}^{n-1} \p_k^2\, G(A,\Delta)
\end{equation}
where $\Delta=(ia)^2 \sum_{k=1}^{n-1} \p_k^2$. The family of realizations \eqref{48}-\eqref{49} includes
special realizations corresponding to the left, right, symmetric left-right and symmetric Weyl orderings for the
enveloping algebra of the Lie algebra \eqref{52}. These realizations are parameterized by
\begin{equation}
\varphi(A)=e^{-A}, \quad \varphi(A)=1, \quad \varphi(A)=e^{-A/2}\quad \text{and}\quad \varphi(A)=A/(e^A-1),
\end{equation}
respectively. We remark that only the symmetric Weyl realization is covariant under the full algebra $so(n)$.

For a given parameter function $\varphi$ and an arbitrary choice of $N_1$, $N_2$ and $G$ one can find the one-forms $\xi_k$
satisfying $[\hat d,\hat x_\mu ] =\xi_\mu$. As in the case of the covariant realizations one can express the
commutator $[\xi_\mu,\hat x_\nu]$ in terms of the one-forms $\xi_\mu$ and partial derivatives $\p_\mu$, but
the general expressions are fairly complicated.

In the following we will focus our attention to a subfamily of the noncovariant realizations which lead to some
interesting results. These realizations are parameterized by $\varphi(A)=e^{-cA}$, $c\in \R$:
\begin{align}
\hat x_k &= x_k\, e^{-cA}, \quad k=1,2,\ldots ,n-1, \\
\hat x_n &= x_n +ia(1-c) \sum_{k=1}^{n-1} x_k\, \p_k.
\end{align}
They include the left, right and symmetric left-right realizations for $c=1$, $c=0$ and $c=1/2$, respectively.
Let us define the exterior derivative by
\begin{equation}
\hat d= \sum_{k=1}^{n-1} dx_k\, \p_k\, e^{(c-1)A}+dx_n\, \p_n
\end{equation}
($N_1=e^{(c-1)A}$, $N_2=1$, $G=0$). Then the corresponding one-forms are given by
\begin{align}
\xi_k &= [\hat d, \hat x_k] = dx_k\, e^{-A}, \quad k=1,2,\ldots ,n-1, \\
\xi_n &= [\hat d,\hat x_n]= dx_n.
\end{align}
The algebra generated by $\hat x_\mu$ and $\xi_\mu$ satisfies the commutation relations
\begin{alignat}{2}
[\xi_k, \hat x_l] &=0, \quad & [\xi_k,\hat x_n] &= -ia\xi_k,   \label{50} \\
[\xi_n,\hat x_l] &=0, \quad & [\xi_n,\hat x_n] &= 0.  \label{51}
\end{alignat}
This algebra satisfies the graded Jacobi relations \eqref{80}.
We note that the relations \eqref{50}-\eqref{51} correspond to the algebra found by Kim et. al. \cite{Kim}
where the commutators are defined in terms of the star-product, except that in our work $\xi_\mu$ and $\xi_\nu$ anticommute.
In particular, for $c=0$ the exterior derivative becomes
\begin{equation}
\hat d = \sum_{k=1}^{n-1} dx_k\, \p_k\, e^{-A}+dx_n\, \p_n = \sum_{\a=1}^n \xi_\a\, \p_\a,
\end{equation}
which is the type II realization of $\hat d$. In addition to the examples in section 3
the commutators \eqref{50}-\eqref{51} also close in the space of one-forms $\xi_\mu$ alone.
Moreover, the right realization ($c=0$) is an example of a type II realization with closed commutator.

The above construction can be extended to any parameter function $\varphi$. It can be shown that for a given
$\varphi$ one can find $N_1$, $N_2$ and $G$ such that $\hat d = \sum_{\a} \xi_\a \p_\a$
and $[\hat d,\hat x_\mu]=\xi_\mu$. However, this may be very complicated as already seen in the natural
realization in section 3.

\section{Extended star-product}

Regarding functions as zero-forms we want to extend the star-product to differential forms of arbitrary degree.
The star-product of differential forms in the context of deformation quantization has been investigated recently
in \cite{Zumino}. The construction of the star-product presented here is valid for a general Lie algebra type
noncommutative space. We recall that the realization of NC coordinates $\hat x_\mu$ in terms of $x_\mu$ and
$\p_\mu$ is given by Eq. \eqref{03}. Also, since the matrix $[\phi_{\mu\nu}]$ is invertible the commutative coordinates
$x_\mu$ admit realization in terms of $\hat x_\mu$ and $\p_\mu$ via Eq. \eqref{59}. The duality between $\hat x_\mu$ and $x_\mu$
induces a vector space isomorphism $\Omega_\phi \colon \mathcal{U}(\g)\to \mathcal{S}$ between the enveloping algebra
$\mathcal{U}(\g)$ of the Lie algebra $\eqref{01}$ and the symmetric algebra $\mathcal{S}$ generated by $x_\mu$, $\mu=
1,2,\ldots ,n$. The isomorphism $\Omega_\phi$ depends on the realization $\phi$, and is given as follows.
Let $1$ denote the unit in $\mathcal{S}$ ($\mathcal{S}$ is isomorphic to the Fock space built on the
vacuum vector $|0\rangle \equiv 1$). Then $x_\mu$ and $\p_\mu$ act on $f\in \mathcal{S}$ in a natural way by
$x_\mu \cdot f = x_\mu f$ and $\p_\mu \cdot f = \frac{\p f}{\p x_\mu}$. In particular,
\begin{equation}\label{75}
x_\mu \cdot 1 = x_\mu, \qquad \p_\mu\cdot 1 = 0.
\end{equation}
For a monomial $\hat f(\hat x)\in \mathcal{U}(\g)$ we define
\begin{equation}
\Omega_\phi \big(\hat f(\hat x)\big) = \hat f(\hat x)\cdot 1 \equiv f(x),
\end{equation}
and extend $\Omega_\phi$ linearly to $\mathcal{U}(\g)$. The map $\Omega_\phi$ is evaluated at $\hat f(\hat x)$
by using the realization \eqref{03} and action \eqref{75}. For example,
\begin{equation}
\Omega_\phi (\hat x_\mu) = \sum_\a \left(x_\a \phi_{\a\mu}(\p)\right)\cdot 1 = x_\mu
\end{equation}
since $\phi_{\a\mu}(\p) = \delta_{\a\mu}+o(\p)$. Similarly, for monomials of order two we have
\begin{equation}
\Omega_\phi (\hat x_\mu \hat x_\nu) = x_\mu x_\nu +\sum_\a x_\a \frac{\p \phi_{\a\mu}}{\p \p_\nu}\cdot 1
\end{equation}
where $\frac{\p \phi_{\a\mu}}{\p \p_\nu}\cdot 1$ is a first-order coefficient in the Taylor expansion of $\phi_{\a\mu}(\p)$.
In general, $\Omega_\phi (\hat x_{\mu_1}\hat x_{\mu_2}\ldots \hat x_{\mu_m})$ is a polynomial in the variables
$x_{\mu_1},x_{\mu_2},\ldots ,x_{\mu_m}$ whose coefficients are given by the Taylor expansion of $\phi_{\mu\nu}$. The computation of
$\Omega_\phi (\hat x_{\mu_1}\hat x_{\mu_2}\ldots \hat x_{\mu_m})$ can be done using a recursive formula. Suppose that
\begin{equation}
\Omega_\phi (\hat x_{\mu_2}\hat x_{\mu_3}\ldots \hat x_{\mu_m}) = p(x_{\mu_2},x_{\mu_3},\ldots ,x_{\mu_m}).
\end{equation}
Then
\begin{align}
\Omega_\phi (\hat x_{\mu_1}\hat x_{\mu_2}\ldots \hat x_{\mu_m}) &= x_{\mu_1}p(x_{\mu_2},x_{\mu_3},\ldots ,x_{\mu_m})  \notag \\
&+\sum_\a x_\a\, \big[\phi_{\a\mu_1},p(x_{\mu_2},x_{\mu_3},\ldots ,x_{\mu_m})\big]\cdot 1.
\end{align}
The commutator in the above expression is calculated according to
\begin{equation}
[\phi_{\a\mu},x_1x_2\ldots x_k] = [\phi_{\a\mu},x_1]x_2\ldots x_k
+ x_1 [\phi_{\a\mu},x_2]\ldots x_k + \cdots + x_1\ldots x_{k-1} [\phi_{\a\mu},x_k].
\end{equation}

The inverse map $\Omega_\phi^{-1}$ is defined analogously. Let $\hat 1$ be the unit in $\mathcal{U}(\g)$.
Define the action of $\hat x_\mu$ on a monomial $\hat f(\hat x)\in \mathcal{U}(\g)$ by
$\hat x_\mu \cdot \hat f(\hat x)=\hat x_\mu \hat f(\hat x)$. The action of $\p_\mu$ on $\hat f(\hat x)$
is defined by $\p_\mu \cdot \hat 1 =0$ and $\p_\mu \cdot \hat f(\hat x) = (\p_\mu \hat f(\hat x))\cdot \hat 1$
where $\p_\mu \hat f(\hat x)$ is expressed using the commutation relations $[\p_\mu, \hat x_\nu]=\phi_{\mu\nu}(\p)$.
For the lowest order vector we have
\begin{equation}\label{76}
\hat x_\mu \cdot \hat 1 = \hat x_\mu, \qquad \p_\mu \cdot \hat 1 =0.
\end{equation}
Then $\Omega_\phi^{-1}$ is given by
\begin{equation}
\Omega_\phi^{-1}\big(f(x)\big) = f(x)\cdot \hat 1 \equiv \hat f(\hat x)
\end{equation}
where $f(x)\cdot \hat 1$ is calculated using the realization \eqref{59} and relations \eqref{76}. For example,
\begin{equation}
\Omega_\phi^{-1}(x_\mu) = \sum_\a \hat x_\a \phi^{-1}_{\a\mu}(\p)\cdot \hat 1 = \hat x_\mu
\end{equation}
since $\phi^{-1}_{\a\mu}(\p) = \delta_{\a\mu}+o(\p)$, and for monomials of order two we have
\begin{equation}\label{77}
\Omega_\phi^{-1} (x_\mu x_\nu) = \hat x_\mu \hat x_\nu +\sum_\a \hat x_\a \frac{\p \phi^{-1}_{\a\mu}}{\p \p_\nu}\cdot \hat 1.
\end{equation}
One can show that the right hand side of Eq. \eqref{77} is invariant under the transposition of indices $\mu \leftrightarrow \nu$,
hence $\Omega_{\phi}^{-1}(x_\mu x_\nu)$ is well-defined. Clearly, $\Omega_\phi$ and $\Omega^{-1}_\phi$ can be readily
extended to $\overline{\mathcal{U}(\g)}$ and $\overline{\mathcal{S}}$, the formal completions of $\mathcal{U}(\g)$
and $\mathcal{S}$. The star-product of $f,g\in \overline{\mathcal{S}}$ is defined by
\begin{equation}\label{56}
(f\star_\phi g)(x) = \big(\hat f(\hat x) \hat g(\hat x)\big)\cdot 1
\end{equation}
where $\hat f(\hat x)=\Omega^{-1}_\phi (f(x))$ and $\hat g(\hat x)=\Omega^{-1}_\phi (g(x))$. In the limit
as the deformation parameter $a\to 0$ the star-product reduces to ordinary product of functions (c.f. Eq. \eqref{55}).
The star-product on the kappa-deformed space was discussed in \cite{Meljanac-1}, \cite{Meljanac-2}, \cite{Meljanac-3};
see also \cite{Vassilevich}.

Equation \eqref{56} defines the star-product of zero-forms. Following the ideas outlined above we want to extend
the star-product to differential forms of arbitrary degree.
Our strategy is to associate to $\omega^k$ a noncommutative form $\hat \omega^k$ such that $\hat \omega^k \cdot 1
= \omega^k$, and define the star-product by
\begin{equation}\label{61}
\omega^k \star_\phi \eta^l = (\hat \omega^k\, \hat \eta^l) \cdot 1.
\end{equation}
It turns out that the star-product \eqref{61} is well-defined provided the commutator $[\xi_\mu,\hat x_\nu]$ is closed
in the space of one-forms $\xi_\mu$ alone. It depends only on the realizations of the coordinates $\hat x_\mu$, hence
we also denote it by $\star_\phi$.

First let us consider the star-product of constant forms.
Recall that the noncommutative one-form $\xi_\mu$ is defined by $\xi_\mu = \sum_\a dx_\a\, h_{\a\mu}(\p)$ where
$h_{\a\mu}$ satisfies Eq. \eqref{10}. The matrix $[h_{\mu\nu}]$ is invertible, hence there is a dual relation
$dx_\mu = \sum_\a \xi_\a\, h^{-1}_{\a\mu}(\p)$. Since $h_{\a\mu}(\p)$ is a power series of the type \eqref{55},
and $dx_\mu$ and $\p_\nu$ commute, we have
\begin{equation}\label{57}
(\xi_{\mu_1}\xi_{\mu_2}\ldots \xi_{\mu_k}) \cdot 1 = dx_{\mu_1}dx_{\mu_2}\ldots dx_{\mu_k}.
\end{equation}
Therefore, to a $k$-form $\omega^k = dx_{\mu_1} dx_{\mu_2}\ldots dx_{\mu_k}$ we associate a unique
noncommutative form $\hat \omega^k=\xi_{\mu_1}\xi_{\mu_2}\ldots \xi_{\mu_k}$ satisfying $\hat \omega^k \cdot 1 = \omega^k$.
The star-product of $\omega^k = dx_{\mu_1}dx_{\mu_2}\ldots dx_{\mu_k}$ and $\eta^l = dx_{\nu_1}dx_{\nu_2}\ldots dx_{\nu_l}$
is trivially given by
\begin{equation}
\omega^k \star_\phi \eta^l = (\xi_{\mu_1}\xi_{\mu_2}\ldots \xi_{\mu_k}\, \xi_{\nu_1}\xi_{\nu_2}\ldots \xi_{\nu_l})\cdot 1.
\end{equation}
In view of Eq. \eqref{57} the star-product of constant forms is undeformed,
\begin{equation}
\omega^k \star_\phi \eta^l = \omega^k\, \eta^l,
\end{equation}
and graded-commutative,
\begin{equation}
\omega^k \star_\phi \eta^l = (-1)^{kl}\, \eta^l\star_\phi \omega^k.
\end{equation}

Now suppose that $\omega^k$ is a general $k$-form $\omega^k = p(x)\, dx_{\sigma_1}dx_{\sigma_2}\ldots dx_{\sigma_k}$ where
$p(x)$ is a monomial in $x_\mu$. Then the associated noncommutative form is given by $\hat \omega^k = \omega^k \cdot \hat 1$
where we define $\xi_\mu \cdot \hat 1 = \xi_\mu$. This yields
\begin{equation}\label{78}
\hat \omega^k = \Omega_\phi^{-1}(p(x)) \xi_{\sigma_1}\xi_{\sigma_2}\ldots \xi_{\sigma_k}.
\end{equation}
Indeed, let us denote $\hat p(\hat x)=\Omega_\phi^{-1}(p(x))$.
Using commutativity of $dx_\mu$ with $x_\mu$ and $\p_\mu$ we obtain
\begin{align}
\hat \omega_k &= \sum_{\rho_1,\ldots ,\rho_k} dx_{\rho_1} dx_{\rho_2}\ldots dx_{\rho_k}\,
\hat p(\hat x)\, h_{\rho_1\sigma_1} h_{\rho_2\sigma_2}\ldots h_{\rho_k\sigma_k} \\
&= dx_{\sigma_1}dx_{\sigma_2}\ldots dx_{\sigma_k}\big(\hat p(\hat x)+o(\p)\big).
\end{align}
Thus,
\begin{equation}
\hat \omega_k \cdot 1 = p(x)\, dx_{\sigma_1}dx_{\sigma_2}\ldots dx_{\sigma_k}=\omega^k
\end{equation}
since $\hat p(\hat x)\cdot 1 = p(x)$. We note that $\hat \omega^k$ given by Eq. \eqref{78} is a
unique noncommutative form (up to reordering of $\hat x_\mu$ in $\hat p(\hat x)$
using the commutation relations \eqref{01}) with the property $\hat \omega^k \cdot 1 =
\omega^k$ in which the NC coordinates are naturally ordered to the left of $\xi_\mu$.
If $\omega^k = p(x) dx_{\mu_1}dx_{\mu_2}\ldots dx_{\mu_k}$ and $\eta^l = q(x) dx_{\nu_1}dx_{\nu_2}\ldots dx_{\nu_l}$,
then Eqs. \eqref{61} and \eqref{78} yield
\begin{equation}\label{79}
\omega^k \star_\phi \eta^l = \big(\hat p(\hat x)\, \xi_{\mu_1}\ldots \xi_{\mu_k}\, \hat q(\hat x)\, \xi_{\nu_1}\ldots \xi_{\nu_l}\big)\cdot 1
\end{equation}
where $\hat p(\hat x)=\Omega_\phi^{-1} (p(x))$ and $\hat q(\hat x)=\Omega_\phi^{-1} (q(x))$.
The star-product \eqref{79} is not graded-commutative since $\hat x_\mu$ and $\xi_\mu$ do not commute.
The product is well-defined provided the commutators $[\xi_\mu,\hat x_\nu]$ are closed in the space
of one-forms $\xi_\mu$. In this case one can use the commutation relations between $\xi_\mu$ and $\hat x_\nu$
to write \eqref{79} in the natural order with $\hat x_\mu$ to the left of $\xi_\mu$ and evalute the star-product
using $\big(\hat p(\hat x)\xi_{\mu_1}\ldots \xi_{\mu_k}\big) \cdot 1 = p(x)dx_{\mu_1}\ldots dx_{\mu_k}$.
In view of earlier considerations, the extended star-product can be defined in the covariant left, right and
noncovariant realizations discussed in sections 3 and 4.
We note that the extended star-product is associative since this property is inherited from associativity of operator
multiplication in the superalgebra $\mathcal{A}$.

Finally, let us consider the exterior derivative acting on the star-product of forms. In the realization of type I the
exterior derivative is undeformed, $\hat d = d \equiv \sum_\a dx_\a \p_\a$. Then one can show that
\begin{equation}
d\omega = (\hat d \hat \omega) \cdot 1,
\end{equation}
where $\hat \omega \cdot 1 = \omega$. Using the star-product \eqref{61} and Leibniz rule \eqref{64} one finds
\begin{equation}
d(\omega\star_\phi \eta) = d\omega \star_\phi \eta + (-1)^{|\omega|}\, \omega \star_\phi d\eta.
\end{equation}
Hence, in type I realization the Leibniz rule for the extended star-product is undeformed. It would
be interesting to invstigate the action of the induced exterior derivative on the star-product of
forms in other realizations when $\hat d$ is given by a general expression \eqref{08}.

\section{Concluding remarks}

In this paper we have investigated the differential algebra of forms on the kappa-deformed space. Our construction of the exterior
derivative $\hat d$ and one-forms $\xi_\mu$ is based on the realizations of NC coordinates $\hat x_\mu$ in terms of
formal power series in the Weyl algebra. We have shown that for each realization of $\hat x_\mu$
there is an infinite family of the exterior derivatives $\hat d$ which uniquely determine the one-forms $\xi_\mu$. The exterior derivative
is a nilpotent operator and it satisfies the undeformed Leibniz rule. The NC coordinates $\hat x_\mu$, derivatives $\p_\mu$
and one-forms $\xi_\mu$ generate a $\mathbb{Z}_2$-graded algebra. The subalgebra generated by $\hat x_\mu$ and $\p_\mu$ is a
deformed Heisenberg algebra. The algebra generated by $\hat x_\mu$ and $\xi_\mu$ is generally not closed under the commutator
bracket since $[\xi_\mu,\hat x_\nu]$ may involve an infininte series in $\p_\mu$.
Only in special cases of the covariant left, right and noncovariant realizations the algebra is closed under the commutator bracket. Furthermore,
the commutator $[\xi_\mu,\hat x_\nu]$ is nonzero in all realizations. For higher-order forms
we have shown that the exterior derivative satisfies the graded Leibniz rule, and the graded Jacobi identity also holds. However,
the graded commutativity law holds only for $\hat x_\mu$-independent forms. In the limit when the deformation parameter $a\to 0$ our theory
reduces to classical results.

The exterior derivative and one-forms have been analyzed in both covariant and noncovariant realizations. In the covariant
case we have found explicit representations of $\hat d$ and $\xi_\mu$ in the left, right and natural realization. We have also found a closed
form expression for the commutator $[\xi_\mu,\hat x_\nu]$ in these realizations, and derived an approximation to first order in $a$
in the natural realization. In the noncovariant case we have constructed a one-parameter family of realizations of $\hat d$ and $\xi_\mu$.
For this family of realizations the commutator $[\xi_\mu,\hat x_\nu]$ is always closed in the space of one-forms $\xi_\mu$.

We have also extended the star-product from zero-forms to differential forms of arbitrary degree. The star-product can be defined
for realizations in which $[\xi_\mu,\hat x_\nu]$ is closed in the space of one-forms $\xi_\mu$. It depends only
on the realizations of both the NC coordinates $\hat x_\mu$. For diffential forms with constant
coefficients the star-product is undeformed and graded-commutative, but for arbitrary forms this is no longer true.
It was shown the the exterior derivative acting on the extended star-product satisfies the undeformed Leibniz rule
in type I realization. It would be interesting to investigate possible relations between our approach to the star-product
of differential forms and the recent work presented in \cite{Zumino}.

Finally, the notion of the twist operator is very important in the construction of the star-product from both the mathematical
(\cite{Drinfeld}, \cite{Majid}) and physical (\cite{Aschieri-2}, \cite{Banerjee}, \cite{Arzano}, \cite{Young}, \cite{Govindarajan})
points of view. The twist operator for zero-forms on the kappa-deformed space was constructed in \cite{Meljanac-4} and \cite{Govindarajan},
and was also considered in \cite{Bu}. However, it remains an open problem to see if there exisits a twist
operator that leads to the star-product of differential forms defined in this work.

\section{Appendix A}

In this appendix we find the solution of the system of equations \eqref{24}-\eqref{25}.
Let us write Eq. \eqref{24} in equivalent forms as
\begin{equation}
(AZ^{-1}-B)\frac{\p H_1}{\p A}+2B\sqrt{1-B}\, \frac{\p H_1}{\p B}+(Z^{-1}-1)H_1 =0.
\end{equation}
We assume that $H_1$ can be factored as $H_1(A,B)=Z F_1(B)$ which leads to the following differential equation for $F_1$,
\begin{equation}\label{26}
2B\sqrt{1-B}\, F_1^\prime (B)+\big(\sqrt{1-B}-1\big) F_1(B)=0.
\end{equation}
The boundary condition for $H_1$ implies that $\lim_{a\to 0}F_1(B)=1$. Now the solution of Eq. \eqref{26}
is readily found to be
\begin{equation}\label{27}
F_1(B) = \frac{2(1-\sqrt{1-B})}{B},
\end{equation}
hence
\begin{equation}
H_1(A,B) = \frac{2(1-\sqrt{1-B})}{B(-A+\sqrt{1-B})}.
\end{equation}

Next, let us consider Eq. \eqref{25} which we write equivalently as
\begin{equation}\label{28}
(AZ^{-1}-B)\frac{\p H_2}{\p A}+2B\sqrt{1-B}\, \frac{\p H_2}{\p B}+\big(2\sqrt{1-B}-1\big) H_2 = -H_1.
\end{equation}
We apply a similar method of ``separation of variables'' assuming that $H_2(A,B)=ZF_2(B)+F_3(B)$.
Inserting the Ansatz for $H_1$ and $H_2$ into Eq. \eqref{28}, and grouping the terms
depending only on $B$ on the right-hand side, we obtain
\begin{multline}\label{29}
AF_2(B)+Z^{-1}\Big(2B\sqrt{1-B}\, F_3^\prime (B)+(2\sqrt{1-B}-1) F_3(B)\Big) = \\
-2B\sqrt{1-B}\, F_2^\prime (B) -\big(2\sqrt{1-B}-1\big) F_2(B)-F_1(B).
\end{multline}
Let us define the function
\begin{equation}
G(B)=2B\sqrt{1-B}\, F_3^\prime (B)+\big(2\sqrt{1-B}-1\big) F_3(B).
\end{equation}
Then the variables in Eq. \eqref{29} can be separated as
\begin{multline}\label{30}
A\big(F_2(B)-G(B)\big) = \\ -2B\sqrt{1-B}\, F_2^\prime (B)-(2\sqrt{1-B}-1)F_2(B)-F_1(B)-\sqrt{1-B}\, G(B).
\end{multline}
We conclude that both sides of the equation must be zero which implies that $F_2$ and $F_3$ satisfy the following system of
differential equations:
\begin{align}
2B\sqrt{1-B}\, F_2^\prime (B)+(3\sqrt{1-B}-1) F_2(B) &= -F_1(B), \label{31} \\
2B\sqrt{1-B}\, F_3^\prime (B) +(2\sqrt{1-B}-1) F_3(B) &= F_2(B).  \label{32}
\end{align}
Using the boundary condition for $H_2$ we find that in the limit $a\to 0$ both $F_2(B)$ and $F_3(B)$ must be finite.
Taking this into account, integration of the system \eqref{31}-\eqref{32} yields
\begin{equation}
F_2(B) = F_3(B) = -2\left(\frac{1-\sqrt{1-B}}{B}\right)^2.
\end{equation}
Therefore,
\begin{equation}
H_2(A,B) = -2(1-A+\sqrt{1-B})\left(\frac{1-\sqrt{1-B}}{B}\right)^2.
\end{equation}

\section{Appendix B}

In this Appendix we give a brief derivation of the result \eqref{40}. We shall do this in two
steps. First we calculate the commutator $[\xi_\mu,\hat x_\nu]$ where $\xi_\mu=\sum_\a dx_\a h_{\a\mu}(\p)$
and $\hat x_\nu$ is given in the natural realization \eqref{20}. We have
\begin{equation}
[\xi_\mu,\hat x_\nu] = Z^{-1} \sum_\a [h_{\a\mu},x_\nu ]\, dx_\a +\p_\nu \sum_\a [h_{\a\mu},iax]\, dx_\a.
\end{equation}
Expressing $h_{\a\mu}$ by Eq. \eqref{41} and making use of
\begin{equation}
\frac{\p f(A,B)}{\p \p_\mu} = i\frac{\p f}{\p A}a_\mu +2a^2 \frac{\p f}{\p B}\p_\mu,
\end{equation}
after some manipulation we find
\begin{equation}\label{45}
\begin{split}
\sum_\a [h_{\a\mu},x_\nu ]dx_\a &= \left(i\frac{\p L_1}{\p A}a_\nu +2a^2 \frac{\p L_1}{\p B}\p_\nu\right) dx_\mu\\
&+\left(iL_3 a_\mu +a^2 L_4\p_\mu\right) dx_\nu +iS_{\mu\nu} (adx)+T_{\mu\nu}(\p dx),
\end{split}
\end{equation}
where we have defined
\begin{multline}
S_{\mu\nu} = L_2\, \delta_{\mu\nu} + 2\left(B\frac{\p L_5}{\p B}+L_5\right) ia_\mu \p_\nu
+\frac{\p L_2}{\p A}\, ia_\nu \p_\mu \\
+ 2a^2 \frac{\p L_2}{\p B}\, \p_\mu \p_\nu-\p^2 \frac{\p L_5}{\p A}\, a_\mu a_\nu,
\end{multline}
\begin{equation}
T_{\mu\nu} = a^2 L_4\, \delta_{\mu\nu} + 2a^2 \frac{\p L_3}{\p B}\, ia_\mu \p_\nu
+a^2 \frac{\p L_4}{\p A}\, ia_\nu \p_\mu
+2a^4 \frac{\p L_4}{\p B}\, \p_\mu \p_\nu - \frac{\p L_3}{\p A}\, a_\mu a_\nu.
\end{equation}
A similar computation yields
\begin{multline}\label{44}
\sum_\a [h_{\a\mu},iax]\, dx_\a \\= a^2 E_1 dx_\mu +(iE_2a_\mu +a^2 E_3 \p_\mu) (iadx)+a^2(iE_4a_\mu + a^2 E_5\p_\mu) (\p dx)
\end{multline}
where the functions $E_i$ are defined by
\begin{align}
E_1 &= 2A\frac{\p L_1}{\p B}-\frac{\p L_1}{\p A},  \\
E_2 &= L_2+L_3+2AL_5+2AB\frac{\p L_5}{\p B}-B\frac{\p L_5}{\p A}, \\
E_3 &= L_4+2A\frac{\p L_2}{\p B}-\frac{\p L_2}{\p A}, \\
E_4 &= L_4+2A\frac{\p L_3}{\p B}-\frac{\p L_3}{\p A}, \\
E_5 &= 2A\frac{\p L_4}{\p B}-\frac{\p L_4}{\p A}.
\end{align}
Combining equations \eqref{45} and \eqref{44} we obtain
\begin{equation}\label{46}
[\xi_\mu,\hat x_\nu] = dx_\mu\, P^{(1)}_\nu + dx_\nu\, P^{(2)}_\mu (\p) + (iadx)\, Q^{(1)}_{\mu\nu} + (\p dx)\, Q^{(2)}_{\mu\nu}.
\end{equation}
where the functions $P_\mu^{(i)}$ and $Q_{\mu\nu}^{(i)}$ are given by
\begin{align}
P^{(1)}_\nu &= Z^{-1}\frac{\p L_1}{\p A}\, ia_\nu + a^2\Big(2Z^{-1}\frac{\p L_1}{\p B}+E_1\Big)\p_\nu, \\
P^{(2)}_\mu &= Z^{-1}L_3\, ia_\mu +a^2 Z^{-1}L_4\, \p_\mu, \\
Q^{(1)}_{\mu\nu} &= Z^{-1}S_{\mu\nu}+E_2\, ia_\mu \p_\nu+a^2 E_3\, \p_\mu \p_\nu, \\
Q^{(2)}_{\mu\nu} &= Z^{-1}T_{\mu\nu}+a^2 E_4\, ia_\mu\p_\nu +a^4 E_5\, \p_\mu \p_\nu.
\end{align}

In the second step we wish to express the commutator \eqref{46} in terms of the one-forms $\xi_\mu$ and derivatives
$\p_\mu$. In order to replace $dx_\mu$ by $\xi_\mu$ we write $dx_\mu =\sum_\a h^{-1}_{\a\mu}(\p)\xi_\mu$ where
$h^{-1}_{\a\mu}$ is the inverse of the matrix $h_{\a\mu}$.
The inverse matrix should have the same index structure as $h_{\a\mu}$, hence we look for $h^{-1}_{\a\mu}$ in the form
\begin{equation}
h^{-1}_{\a\mu}(\p) = G_1 \delta_{\a\mu}+iG_2\, a_\a \p_\mu +iG_3\, a_\mu \p_\a +a^2 G_4\, \p_\a \p_\mu -\p^2 G_5\, a_\a a_\mu.
\end{equation}
The condition $\sum_\a h_{\a\beta}\, h^{-1}_{\beta\mu} = \delta_{\a\mu}$ implies that the functions $G_k$ satisfy the
following system of equations:
\begin{align}
G_1 &= L_1^{-1}, \\
-(L_1+AL_2-BL_5)G_2 - B(L_2+AL_5)G_4 &= L_2 L_1^{-1},  \\
(L_3-AL_4)G_2 - (L_1+AL_3+BL_4)G_4 &= L_4 L_1^{-1}, \\
-(L_1+AL_3+BL_4)G_3 + B(L_3-AL_4)G_5 &= L_3 L_1^{-1}, \\
-(L_2+AL_5)G_3 - (L_1+AL_2-BL_5)G_5 &= L_5 L_1^{-1}.
\end{align}
The solution of the system is given by
\begin{align}
G_2 &= \frac{1}{M}\Big[-(L_1+AL_3+BL_4)L_2+B(L_2+AL_5)L_4\Big],   \\
G_3 &= -\frac{1}{M}\Big[(L_1+AL_2-BL_5)L_3+B(L_3-AL_4)L_5\Big], \\
G_4 &= -\frac{1}{M}\Big[(L_3-AL_4)L_2+(L_1+AL_2-BL_5)L_4\Big], \\
G_5 &= \frac{1}{M}\Big[(L_2+AL_5)L_3-(L_1+AL_3+BL_4)L_5\Big],
\end{align}
where
\begin{equation}
M= L_1\Big[(L_1+AL_2-BL_5)(L_1+AL_3+BL_4)+B(L_2+AL_5)(L_3-AL_4)\Big].
\end{equation}
Now, with the functions $G_k$ defined as above, we have
\begin{align}
dx_\mu &= \sum_\a h^{-1}_{\a\mu}(\p)\, \xi_\a \notag \\
&= G_1 \xi_\mu +\left(\p^2 G_5 ia_\mu + G_2\p_\mu\right) (ia\xi)+\left(G_3 ia_\mu + a^2 G_4 \p_\mu\right) (\p\xi). \label{47}
\end{align}
Using Eq. \eqref{47} to eliminate $dx_\mu$ from the commutator \eqref{46} we obtain
\begin{equation}\label{70}
[\xi_\mu,\hat x_\nu] =\xi_\mu\, \frac{P_\nu^{(2)}}{L_1} + \xi_\nu\, \frac{P_\mu^{(2)}}{L_1} + (ia\xi)\, R^{(1)}_{\mu\nu} + (\p\xi)\, R^{(2)}_{\mu\nu},
\end{equation}
where $R^{(1)}_{\mu\nu}$ and $R^{(2)}_{\mu\nu}$ are defined by
\begin{multline}
R^{(1)}_{\mu\nu} = \p^2 G_5 (P^{(1)}_\nu\, ia_\mu + P^{(2)}_\mu\, ia_\nu)+G_2(P^{(1)}_\nu\, \p_\mu+P^{(2)}_\mu\, \p_\nu) \\
+(G_1+AG_2-BG_5)\, Q^{(1)}_{\mu\nu} +\p^2 (G_2+AG_5)\, Q^{(2)}_{\mu\nu},
\end{multline}
\begin{multline}
R^{(2)}_{\mu\nu} = G_3(P^{(1)}_\nu\, ia_\mu +P^{(2)}_\mu\, ia_\nu)+a^2 G_4 (P^{(1)}_\nu\, \p_\mu +P^{(2)}_\mu\, \p_\nu)  \\
+a^2(AG_4-G_3)\, Q^{(1)}_{\mu\nu} + (G_1+AG_3+BG_4)\, Q^{(2)}_{\mu\nu}.
\end{multline}
Tracing back the computations we can express the commutator \eqref{70} explicitly in terms of $L_1,\ldots ,L_5$
and their partial derivatives, but the expressions are cumbersome and not useful for practical calculations.

\textit{Aknowledgements.} This work is supported by the Croatian Ministry of Science, Education
and Sports grants no. 098-0000000-2865 and 177-0372794-2816.


\end{document}